\documentclass[
showkeys,12pt,
preprint,preprintnumbers,nofootinbib,
groupedaddress,superscriptaddress,amsmath,amssymb]{revtex4}
\usepackage{graphicx}
\usepackage{dcolumn}
\usepackage{bm}
\usepackage{amssymb}
\usepackage{amsmath}
\usepackage{epsfig}    
\usepackage{color}
\usepackage{slashed}
\usepackage{hhline}

\def\be{\begin{equation}}
\def\ee{\end{equation}}
\newcommand{\bea}{\begin{eqnarray}}
\newcommand{\eea}{\end{eqnarray}}
\newcommand{\nn}{\nonumber}

\numberwithin{equation}{section}

\begin{document}

\title{A Three-loop Neutrino Model with Leptoquark Triplet Scalars}
\preprint{KIAS-P17002}
\author{Kingman Cheung}
\email{cheung@phys.nthu.edu.tw}
\affiliation{Physics Division, National Center for Theoretical Sciences, 
Hsinchu, Taiwan 300}
\affiliation{Department of Physics, National Tsing Hua University, 
Hsinchu 300, Taiwan}
\affiliation{Division of Quantum Phases and Devices, School of Physics, 
Konkuk University, Seoul 143-701, Republic of Korea}

\author{Takaaki Nomura}
\email{nomura@kias.re.kr}
\affiliation{School of Physics, KIAS, Seoul 130-722, Korea}

\author{Hiroshi Okada}
\email{macokada3hiroshi@cts.nthu.edu.tw}
\affiliation{Physics Division, National Center for Theoretical Sciences, 
Hsinchu, Taiwan 300}

\date{\today}

\begin{abstract}
We propose a three-loop neutrino mass model with a few leptoquark scalars 
in $SU(2)_L$-triplet form, through which we can explain the anomaly 
of $B\to K^{(*)} \mu^+\mu^-$, a sizable muon $g-2$ and a bosonic dark matter 
candidate, and at the same time satisfying all the constraints 
from lepton flavor violations.
We perform global numerical analyses and show the allowed regions, 
in which we find somewhat restricted parameter space, such as 
the mass of dark matter candidate and various components of the 
Yukawa couplings in the model.
 \end{abstract}
\maketitle

\section{Introduction}

Recently, there was an $2.6\sigma$ anomaly in lepton-universality violation
in the ratio 
$R_K \equiv B(B\to K\mu\mu)/B(B\to K ee) = 0.745 ^{+0.090}_{-0.074} \pm 0.036$ 
by the LHCb Collaboration~\cite{lhcb-2014}. In addition, sizable 
deviations were observed in angular distributions of 
$B \to K^*\mu\mu$ \cite{lhcb-2013}.
The results can be interpreted by a large negative
contribution to the Wilson coefficient $C_9$ of the semileptonic operator
$O_9$, and also contributions to other Wilson coefficients, in particular to 
$C'_9$ \cite{Descotes-Genon:2015uva, Hiller:2014yaa, Hiller:2016kry, Descotes-Genon:2013wba}.

The discrepancy between the theoretical prediction and experimental value on
the muon anomalous magnetic dipole moment has been a long-standing 
problem, which stands at $3.6\sigma$ level with the deviation from the 
SM prediction at \cite{pdg}. 
\begin{equation*}
 \Delta a_\mu  = a_\mu^{\rm exp}  - a_\mu^{\rm SM} = 288(63)(49) \times 10^{-11}. 
\end{equation*}
If one insists on fulfilling the muon $g-2$ within $1\sigma-2\sigma$ of the
experimental value in any models, it puts a strong constraint on the
parameter space. For example, it requires a relatively light spectrum
in the supersymmetric particles in the MSSM  in order to bring the 
prediction to be within $1\sigma-2\sigma$ of the experimental value.
A number of leptoquark models have been proposed to solve the
$B\to K^{(*)}\mu\mu$ anomaly, but however it is very hard to satisfy simultaneously
the muon $g-2$: see for example Ref.~\cite{Cheung:2016fjo}.

In this work, we propose a three-loop neutrino mass model with a 
few leptoquark scalars in $SU(2)_L$-triplet form.
We attempt to use the model to explain the anomaly 
of $B^{(*)}\to K\mu^+\mu^-$, to achieve a sizable muon $g-2$, and 
to provide a bosonic dark matter candidate, and at the same time 
satisfying all the constraints from lepton flavor violations.
The concrete model is based on the SM symmetry and a $Z_2$ symmetry as 
$SU(3)_C \times SU(2)_L \times U(1)_Y \times Z_2$. The model consists 
of the SM fields, 3 additional leptoquark triplet fields $\Delta_{1,2,3}^a$, 
and one colorless doublet scalar field $\eta$.  These fields are assigned
different $Z_2$ {parities} and hypercharges in such a way that 
each of the Yukawa-type couplings contributes to either neutrino mass, 
$B \to K^{(*)}\mu\mu$ anomaly, muon $g-2$, or the dark matter interactions.
In this way, although the model contains more parameter, it can however 
explain all the above anomalies. The achievements of the model can be
summarized in the following.
\begin{enumerate}
\item
 The neutrino mass pattern and oscillation can be accommodated with the
Yukawa coupling terms $f,g,h$ in three-loop diagrams~\footnote{See refs.~\cite{Krauss:2002px, Aoki:2008av, Gustafsson:2012vj} for representative three loop neutrino mass models}.

\item 
The Yukawa coupling term $f$ can give useful contributions to the Wilson
coefficients $C_{9,10}$ in such a way that it can explain successfully
the $B\to K^{(*)}\mu\mu$ anomaly.

\item
 The muon $g-2$ receives a large contribution from the Yukawa coupling term 
$r$. With some adjustment of the parameters a level of $10^{-9}$ is 
possible.

\item
 It provides a dark matter (DM) candidate $\eta_R$, the real part of the 
neutral component of the $\eta$ field with correct relic density.

\item
 The model can satisfy all the existing constraints from the lepton-flavor
violations (LFVs), meson mixings, and rare $B$ decays.
\end{enumerate}

This paper is organized as follows.
In Sec.~II, we describe the neutrino mass matrix
and the solution to the anomaly in $b\to s \mu \bar \mu$.
In Sec.~III, we discuss various constraints of the model, including
lepton-flavor violations, FCNC's, oblique parameters, and dark matter. 
In Sec. IV, we present the numerical analysis and allowed parameter space,
followed by the discussion on collider phenomenology.
Sec.~IV is devoted for conclusions and discussion.

\section{The Model}

In this section, we describe the model setup, derive the formulas 
for the active neutrino mass matrix, and calculate the contributions
to $b\to s \mu \bar \mu$.

 \begin{widetext}
\begin{center} 
\begin{table}
\begin{tabular}{|c||c|c|c||c|c||c|}\hline\hline  
&\multicolumn{3}{c||}{Quarks} & \multicolumn{2}{c||}{Leptons}& \multicolumn{1}{c|}{Vector Fermions} \\\hline
& ~$Q_{L_i}^a$~ & ~$u_{R_i}^a$~ & ~$d_{R_i}^a$ ~ 
& ~$L_{L_i}$~ & ~$e_{R_i}$ ~ & ~$L'_{i}$ 
\\\hline 
$SU(3)_C$ & $\bm{3}$  & $\bm{3}$  & $\bm{3}$  & $\bm{1}$& $\bm{1}$& $\bm{1}$   \\\hline 
$SU(2)_L$ & $\bm{2}$  & $\bm{1}$  & $\bm{1}$  & $\bm{2}$& $\bm{1}$& $\bm{2}$   \\\hline 
$U(1)_Y$ & $\frac16$ & $\frac23$  & $-\frac{1}{3}$  & $-\frac12$  & $-1$  & $-\frac12$\\\hline
$Z_2$ & $+$ & $+$  & $+$ & $+$ & $+$ & $-$  \\ \hline
\end{tabular}
\caption{Field contents of fermions
and their charge assignments under 
$SU(3)_C\times SU(2)_L\times U(1)_Y\times Z_2$, where the 
superscript (subscript) index $a\,(i)=1-3$ represents the color (flavor). }
\label{tab:1}
\end{table}
\end{center}
\end{widetext}
\begin{table}
\centering {\fontsize{10}{12}
\begin{tabular}{|c||c|c||c|c|c|}\hline\hline
&~ $\Phi$  ~&~ $\eta$  ~&~ $\Delta_1^a$~&~ $\Delta_2^a$~&~ $\Delta_3^a$ \\\hline
$SU(3)_C$ & $\bm{1}$  & ${\bm{ 1}}$ & $\bm{3}$  & $\bar{\bm{ 3}}$ & $\bar{\bm{3}}$ \\\hline 
$SU(2)_L$ & $\bm{2}$   & $\bm{2}$  & $\bm{3}$  & ${\bm{ 3}}$ & $\bm{3}$  \\\hline 
$U(1)_Y$ & $\frac12$  & $\frac{1}{2}$ & $\frac{2}{3}$   & $\frac{1}{3}$ & $\frac{1}{3}$  \\\hline
$Z_2$ & $+$   & $-$  & $-$ & $-$  & $+$ \\\hline
\end{tabular}%
} 
\caption{Field contents of bosons
and their charge assignments under  
$SU(3)_C\times SU(2)_L\times U(1)_Y\times Z_2$, 
where the superscript index $a=1-3$ represents the color. }
\label{tab:2}
\end{table}
\subsection{ Model setup}
We show all the field contents and their charge assignments in 
Table~\ref{tab:1} for the fermionic sector and  in Table~\ref{tab:2} for the 
bosonic sector.~\footnote{{The same contents of the field are found in the systematic analysis in the last part of Table 3 of ref.~\cite{Chen:2014ska}.}} 
Under this framework, the relevant part of the renormalizable Lagrangian 
and Higgs potential related to the neutrino masses are given by
\begin{align}
\label{eq:Yukawa}
-{\cal L}^{}&=
y_{\ell_{i}}\bar L_{L_i }\Phi e_{R_i} 
+ f_{ij}\bar L_{L_i} \Delta^{\dag}_3 (i\sigma_2) Q^c_{L_j}
+g_{ij} \bar L'_{R_i} \Delta^\dag_1 Q_{L_j} 
+h_{ij} \bar L'_{L_i} \Delta^\dag_2 Q^c_{L_j} 
+r_{ij} \bar L'_{L_i} \eta e_{R_j}\nn\\
&+ M_i \bar L'_{L_i} L'_{R_i}
-\lambda_0 \eta^\dag \Delta_3 \Delta_1 \Phi^* -\lambda'_0 \eta^\dag \Delta_3 \Delta_2^* \Phi
-\lambda_5(\eta^\dag \Phi)^2
+{\rm h.c.},
\end{align}
where we have defined $L'\equiv [N,E]^T$, $\sigma_2$ is the second 
Pauli matrix and we have abbreviated the trivial terms for the Higgs potential.
The scalar fields can be parameterized as 
\begin{align}
&\Phi =\left[
\begin{array}{c}
0\\
\frac{v+\phi}{\sqrt2}
\end{array}\right],\quad 
\eta =\left[
\begin{array}{c}
\eta^+\\
\frac{\eta_R+i\eta_I}{\sqrt2}
\end{array}\right],\quad 
\Delta_1 =\left[
\begin{array}{cc}
\frac{\delta_{2/3}^{(1)}}{\sqrt2} & \delta_{5/3}^{(1)} \\
\delta_{-1/3}^{(1)} & -\frac{\delta_{2/3}^{(1)}}{\sqrt2}
\end{array}\right],\nn\\
&
\Delta_2 =\left[
\begin{array}{cc}
\frac{\delta_{1/3}^{(2)}}{\sqrt2} & \delta_{4/3}^{(2)} \\
\delta_{-2/3}^{(2)} & -\frac{\delta_{1/3}^{(2)}}{\sqrt2}
\end{array}\right],\quad
\Delta_3 =\left[
\begin{array}{cc}
\frac{\delta_{1/3}^{(3)}}{\sqrt2} & \delta_{4/3}^{(3)} \\
\delta_{-2/3}^{(3)} & -\frac{\delta_{1/3}^{(3)}}{\sqrt2}
\end{array}\right],
\label{component}
\end{align}
where the subscript next to the each field represents the electric charge
of the field, $v= 246$ GeV, 
and $\Phi$ is written {in the form after the Goldstone fields are}
aboserbed as the longitudinal components of $W$ and $Z$ bosons.
{Notice here that each of the components of $\Delta_3$ and $\eta$ is in mass eigenstate, since there are no mixing terms 
that are assured by the $Z_2$ and $U(1)_Y$ symmetries. 
On the other hand, components of $\Delta_1$ and $\Delta_{2}$ can mix via $\Phi^* \Phi^* \Delta_1 \Delta_2$ term. 
In the following analysis, we ignore such mixing effects 
assuming the relevant coupling is small. }

{\it  Oblique parameters}: Each of the mass components among $\Delta_i$ 
is strongly restricted by the oblique parameters. 
In order to evade such a strong constraint, we simply assume that 
each of the components should be of the same mass~\cite{Cheung:2016frv}.
Thus, we define $m_{\Delta_i}$ as the mass for the components of $\Delta_i$. 
On the other hand, each component of $\eta$ cannot have the same mass, 
because the neutrino mass is proportional to the mass difference 
between the components of $\eta$, as you shall see later. 
Hence, we consider the oblique parameter constraints on $\eta$, 
which are characterized by $\Delta T$ and $\Delta S$. 
Their formulae are given by~\cite{Barbieri:2006dq}
\begin{align}
\Delta T&=\frac{F[\eta^\pm,\eta_I]+F[\eta^\pm,\eta_R]-F[\eta_I,\eta_R]}{32\pi^2 \alpha_{em} v^2}, \
\Delta S=\frac{1}{2\pi} \int_0^1x(1-x) \ln\left[\frac{x m_{\eta_R}^2 + (1-x) m_{\eta_I}^2}{m_{\eta^\pm}^2}\right],
\end{align}
where $\alpha_{em}\approx 1/137$ is the fine structure constant, and
\begin{align}
F[a,b]= \frac{m_a^2+m_b^2}{2}-\frac{m_a^2m_b^2}{m_a^2-m_b^2}\ln\left[\frac{m_a^2}{m_b^2}\right], \ m_a\neq m_b.
\end{align}
The experimental bounds are given by \cite{pdg}
\begin{align}
(0.05 - 0.09) \le \Delta S \le (0.05 + 0.09), \quad (0.08 - 0.07) \le \Delta T \le (0.08 + 0.07).
 \end{align}
We consider these constraints in the numerical analysis.

\begin{figure}[tb]
\begin{center}
\includegraphics[width=65mm]{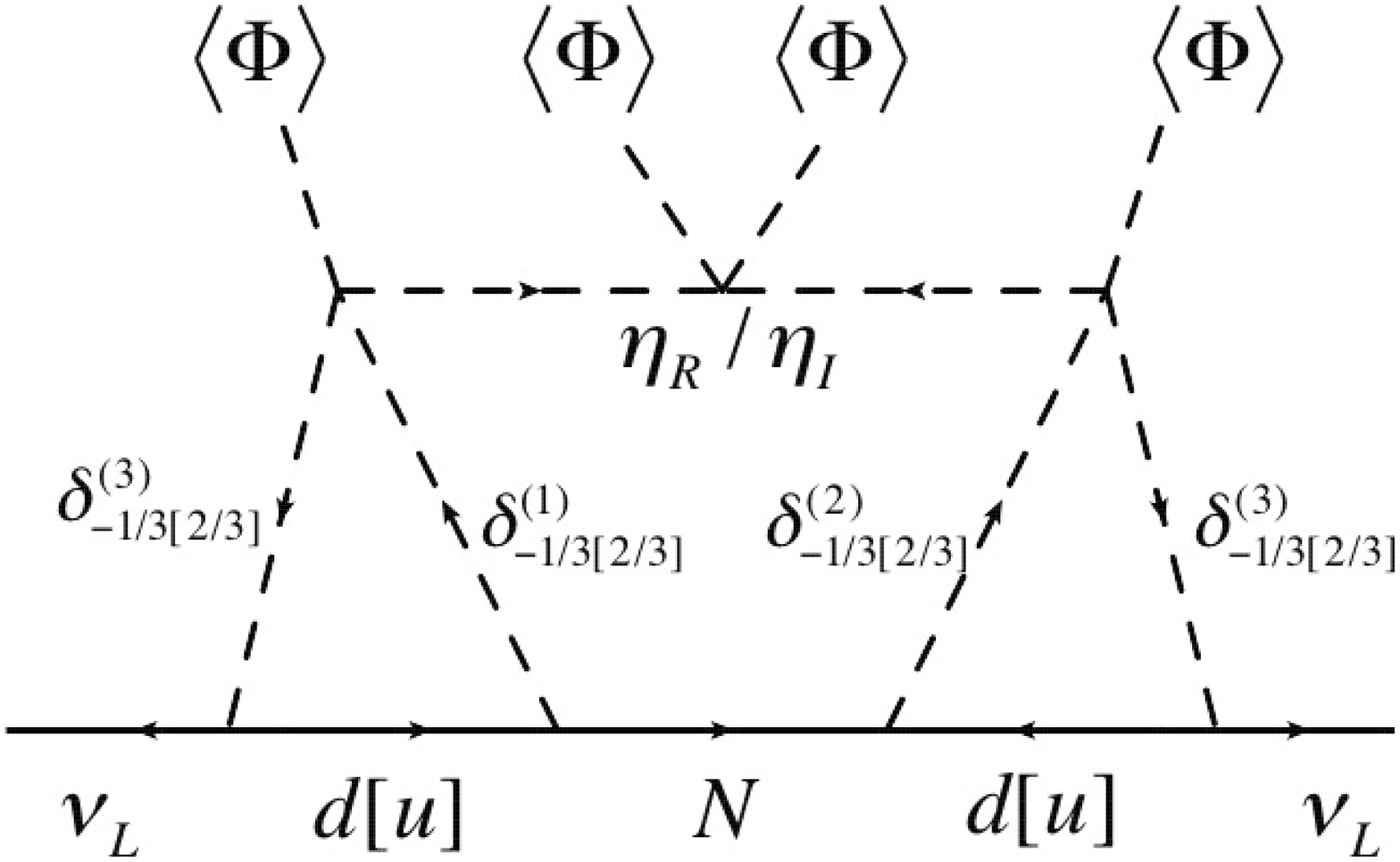}
\caption{
Neutrino mass matrix at the three-loop level, where we have two kind of diagrams that are running up-quarks and down quarks inside the loop. 
}
\label{fig:neut}
\end{center}
\end{figure}

{\it Active neutrino mass matrix}:
The neutrino mass matrix is induced at three-loop level as shown 
in Fig.~\ref{fig:neut}, and its formula 
is generally given by 
\begin{align}
{\cal M}_{\nu_{ij}}&={\cal M}_{\nu_{ij}}^d + {\cal M}_{\nu_{ij}}^u +  tr.
,\ [ {\cal M}_{\nu}^u=2 {\cal M}_{\nu}^d(d\to u,\delta_{1/3}^{(i)}\to \delta_{2/3}^{(i)})],\\
{\cal M}_{\nu_{ab}}^d&=
\frac{3^2\lambda_0\lambda'_0(m_R^2 - m_I^2)v^2}{2\sqrt2(4\pi)^6 M^4_{max}} \sum_{(a,b,c)=1}^3 
f_{ia}g^T_{ab} M_b h^*_{bc} f^T_{cj} 
F_{III}[r_{\Delta_1},r_{\Delta_2} ,  r_{\Delta_3}, r_{b}, r_{R},r_{I},r_{d_c}, r_{d_a}],
\end{align}
where we used the shorthand notation $m_{R/I}\equiv m_{\eta_{R/I}}$, and 
define $M_{\rm Max}\equiv$ Max[$M_b,m_{\Delta_i}, m_{R}, m_I$], 
$r_f\equiv m_f^2/M_{\rm Max}^2$, and the three-loop function $F_{III}$ 
is given in the Appendix.
Here we adopt an assumption $M_{max}=m_{\Delta_3}$, and require 
$1{\rm TeV}\lesssim m_{\Delta_i}$ (which suggests $x_d,x_u\approx0$), which 
is {required by the direct bound on leptoquarks~\cite{Cheung:2016frv}. }
In this case, the neutrino mass matrix can be simplified as
\begin{align}
{\cal M}_{\nu_{ij}}&\approx
\frac{3^3\lambda_0\lambda'_0(m_R^2 - m_I^2)v^2}{2\sqrt2(4\pi)^6 m^4_{\Delta_3}}
\left[f g^T (M F_{III}) h^*  f^T\right]_{ij} + tr.,
\end{align}
 where we have abbreviated the symbol of summation and the 
argument of $F_{III}$.
Then we derive the Yukawa coupling in terms of the experimental values 
and the parameters by introducing an arbitrary anti-symmetric matrix 
with complex values $A$~\cite{Okada:2015vwh}, that is  $A ^T +A=0$, as follows:
 \begin{align}
& g=\frac12 R^{-1} (h^\dag)^{-1}\left[f^{-1} V_{MNS} D_\nu V_{MNS}^T (f^T)^{-1} + A \right]^T,\\
& {\rm or}\nn\\
 & h=\frac12 R^{*-1} (g^\dag)^{-1}\left[f^{-1} V_{MNS} D_\nu V_{MNS}^T (f^T)^{-1} + A \right]^*,
  \end{align}
 where we shall adopt the former formula in the numerical analysis below, 
and we define $D_\nu\equiv V_{MNS}^T  {\cal M}_\nu V_{MNS}$ and  parametrize as 
\begin{align}
 & R=\frac{3^3\lambda_0\lambda'_0(m_R^2 - m_I^2)v^2 MF_{III}}{2\sqrt2(4\pi)^6 m^4_{\Delta_3}},\quad
A\equiv
 \left[\begin{array}{ccc} 
0 &a_{12} & a_{13} \\
-a_{12} & 0 & a_{23} \\
-a_{13} & -a_{23} & 0 \\
  \end{array}
\right].
\end{align}
Here we assume one massless neutrino (with normal ordering) in the 
numerical analysis below.

{\it On the term $f$}:
The new physics contributions to account for the $B\to K^{(*)}\mu\mu$ 
anomaly~\cite{lhcb-2013} can be interpreted as 
the shifts in the Wilson coefficients $C_{9,10}$.
In our model, the relevant Wilson coefficients can be calculated
as follows~\cite{Cheung:2016frv}:
\begin{align}
(C_9)^{\mu \mu}&=-(C_{10} )^{\mu\mu}=-\frac{1}{C_{\rm SM}}
\frac{f_{b\mu} f_{s\mu} }{4 m_{\Delta_3}^2},\quad
C_{\rm SM}\equiv \frac{V_{tb} V^*_{ts} G_F\alpha_{\rm em}}{\sqrt2 \pi},
\end{align}
where $m_{\Delta_3}\equiv m_{\delta_{4/3}^{(3)}}$, ${\rm G_F}\approx1.17\times 10^{-5}$ GeV$^{-2}$ is the Fermi constant.
We can then compare them to 
the {\it best-fit values of $C_{9,10}$} from a global analysis based on the 
LHCb data in 
Ref.~\cite{Descotes-Genon:2015uva} as
\begin{align}
C_{9}=-C_{10}: \ -0.68\; . 
\label{eq:Cp910}
\end{align}
Here we also have to work within the $ -0.75 \lesssim C_9 \lesssim -0.35$ 
in order to satisfy the  the LHCb measurement of 
$R_K = B(B^+ \to K^+ \mu^+\mu^-)/B(B^+ \to K^+ e^+e^-)=0.745^{+0.090}_{-0.074} 
\pm 0.036$, which shows a $2.6\sigma$ deviation from the SM 
prediction~\cite{Cheung:2016frv}.
{\it Notice here that
various constraints arising from the $f$ term include 
$B_{d/s}\to \ell^+\ell^-$, $\ell_i\to \ell_j\gamma$, which were given in
Refs.~\cite{Cheung:2016frv} and~\cite{Cheung:2016fjo}, and we consider 
these constraints in the current numerical analysis. 
Although the muon $g-2$ is also induced from this term, the typical 
order is $10^{-12}\sim 10^{-13}$ with a negative sign~\cite{Cheung:2016frv}. 
Thus, we neglect this contribution to the muon $g-2$.}

 {\it On the terms $g$ and $h$}:
 The main constraint on $g$ and $h$ comes from {$ B(b\to s\gamma)$}.
The partial decay width for $b\to s\gamma$ is given by
\begin{align}
&\Gamma(b\to s\gamma)\approx \frac{3 \alpha_{em} m_b^5}{4(4\pi)^4}
\left|
\frac{g^\dag_{2a} g_{a3}}2 F_{bs\gamma}[\delta_{2/3}^{(1)},a]-
h^\dag_{3a}h_{a2}\left(\frac53 F_{bs\gamma}[a,\delta_{1/3}^{(2)}] +  F_{bs\gamma}[\delta_{1/3}^{(2)},a] \right)
\right|^2,\\
&
F_{bs\gamma}[a,b]=\frac{2 m_a^6+3 m_a^4 m_b^2-6 m_a^2 m_b^4+ m_b^6+ 12 m_a^4 m_b^2\ln[m_b/m_a]}{12(m_a^2-m_b^2)^4},
\end{align}
 then the branching ratio and its experimental bound~\cite{Lees:2012wg} are
given by 
 \begin{align}
 B(b\to s\gamma)\approx
&\frac{\Gamma(b\to s\gamma)}{\Gamma_{tot.}} \lesssim 3.29\times 10^{-4}\;,
\end{align}
 where  $\Gamma_{tot.}\approx 4.02\times10^{-13}$ GeV is the 
total decay width of the bottom quark. In our numerical analysis, we consider this constraint only for the $g$ and $h$ terms.

{\it On the term $r$:}
This term is very important in our model because it can induce a large
contribution to the muon $g-2$ and explain the relic density of dark matter (DM)
if we assume the $\eta_R$ to be the DM candidate.
First of all, let us consider the LFVs processes,
$\ell_a\to\ell_b\gamma$, via one-loop diagrams. 
The branching ratio is given by
\begin{align}
 B(\ell_a\to\ell_b \gamma)
=
\frac{48\pi^3 C_{ab}\alpha_{\rm em}}{{\rm G_F^2} m_a^2 }(|(a_R)_{ab}|^2+|(a_L)_{ab}|^2),
\end{align}
where $m_{a(b)}$ is the mass for the charged-lepton eigenstate, 
$C_{ab}\approx(1, 0.1784,0.1736)$ for $(a,b)=(2,1), (3,1), (3,2)$, and  $a_{L(R)}$  is simply given by
\begin{align}
&(a_{L(R)})_{ab} \approx
 \sum_{i=1}^3\frac{ r^\dag_{bi} r_{ia} m_a}{(4\pi)^2}
\left(F_{lfv}^{L(R)}[N_i,\eta^\pm]-\frac12 F_{lfv}^{L(R)}[\eta_{I},E_i]-\frac12 F_{lfv}^{L(R)}[\eta_{R},E_i]\right), \\
&F_{lfv}^{L}[a,b]=\frac{m_a^4+2 m_a^2m_b^2+2 m_b^2(m_a^2+m_b^2)\ln\left(\frac{m_b^2}{m_a^2+m_b^2}\right)}{6 m_a^6},\
F_{lfv}^{R}[a,b]=\frac{m_a^2+m_b^2 \ln\left(\frac{m_b^2}{m_a^2+m_b^2}\right)}{6 m_a^4},
\label{eq:g-2}
\end{align} 
where {the mass of $E(N)_a$ is defined by $M_{E(N)_a}$}.
Current experimental upper bounds are given 
by~\cite{TheMEG:2016wtm, Adam:2013mnn}
  \begin{align}
  B(\mu\rightarrow e\gamma) &\leq4.2\times10^{-13},\quad
B(\tau\rightarrow \mu\gamma)\leq4.4\times10^{-8},
\quad   B(\tau\rightarrow e\gamma) \leq3.3\times10^{-8}~.
 \label{expLFV}
 \end{align}
 
 {\it Muon $g-2$}:
The muon anomalous magnetic moment is simply given by
$\Delta a_\mu\approx -m_\mu[a_L+a_R]_{22}$ in  Eq.~(\ref{eq:g-2}).
Experimentally, it has been measured with 
a high precision, and its deviation from the SM prediction is 
$\Delta a_\mu={\cal O}(10^{-9})$~\cite{g-2}. 
{It would be worth mentioning a new contribution to the 
leptonic decay of the $Z$ boson. In our case, the $Z$ boson can decay 
into {a pair of charged leptons with a correction 
at one-loop level,}
and it is proportional to the Yukawa couplings related to the 
muon $g-2$. Therefore it can be enhanced due to the large Yukawa couplings.
However, we have checked that this mode is within the experimental 
bound: $B(Z\to \ell\bar\ell)\lesssim 3\times 10^{-2}$. 
}

 {\it $Q-\bar Q$ mixing}:   
The forms of $K^0-\bar K^0$, $B_d^0-\bar B_d^0$, and $D^0-\bar D^0$ mixings are, respectively, given by
{\begin{align}
\Delta m_K&\approx
\frac{1}{(4\pi)^2}
\sum_{i,j=1}^3
\left[g_{i2} g^\dag_{1i}g^\dag_{1j} g_{j2} \left(F^K_{box}[N_i,N_j,\delta_{1/3}^{(1)}]+\frac{F^K_{box}[E_i,E_j,\delta_{2/3}^{(1)}]}4\right)\right.
\nn\\&+
\left.f_{i1}^\dag f_{2i}f_{2j} f^\dag_{j1}\left(F^K_{box}[\ell_i,\ell_j,\delta_{4/3}^{(3)}]+\frac{F^K_{box}[\nu_i,\nu_j,\delta_{1/3}]}4\right)
\right.
\nn\\&+
\left.
h_{i1}^\dag h_{2i}h_{2j} h^\dag_{j1}\left(F^K_{box}[E_i,E_j,\delta_{4/3}^{(2)}]+\frac{F^K_{box}[N_i,N_j,\delta_{1/3}^{(2)}]}4\right) \right]
\lesssim 3.48\times10^{-15}[{\rm GeV}],\label{eq:kk}\\
\Delta m_{B_d}&\approx
\frac{1}{(4\pi)^2}
\sum_{i,j=1}^3
\left[g_{i3} g^\dag_{1i}g^\dag_{1j} g_{j3} \left(F^B_{box}[N_i,N_j,\delta_{1/3}^{(1)}]+\frac{F^B_{box}[E_i,E_j,\delta_{2/3}^{(1)}]}4\right)\right.
\nn\\&+
\left.f_{i3}^\dag f_{1i}f_{1j} f^\dag_{j3}\left(F^B_{box}[\ell_i,\ell_j,\delta_{4/3}^{(3)}]+\frac{F^B_{box}[\nu_i,\nu_j,\delta_{1/3}]}4\right)
\right.
\nn\\&+
\left.
h_{i3}^\dag h_{1i}h_{1j} h^\dag_{j3}\left(F^B_{box}[E_i,E_j,\delta_{4/3}^{(2)}]+\frac{F^B_{box}[N_i,N_j,\delta_{1/3}^{(2)}]}4\right) \right]
 \lesssim 3.36\times10^{-13} [{\rm GeV}], \\
\Delta m_D&\approx
\frac{1}{(4\pi)^2}
\sum_{i,j=1}^3
\left[g_{i2} g^\dag_{1i}g^\dag_{1j} g_{j2} \left(F^D_{box}[E_i,E_j,\delta_{5/3}^{(1)}]+\frac{F^D_{box}[N_i,N_j,\delta_{2/3}^{(1)}]}4\right)\right.
\nn\\&+
\left.f_{i1}^\dag f_{2i}f_{2j} f^\dag_{j1}\left(F^D_{box}[\ell_i,\ell_j,\delta_{4/3}^{(3)}]+\frac{F^D_{box}[\nu_i,\nu_j,\delta_{1/3}]}4\right)
\right.
\nn\\&+
\left.
h_{i1}^\dag h_{2i}h_{2j} h^\dag_{j1}\left(F^D_{box}[N_i,N_j,\delta_{2/3}^{(2)}]+\frac{F^D_{box}[E_i,E_j,\delta_{1/3}^{(2)}]}4\right) \right]
 \lesssim 6.25\times10^{-15}[{\rm GeV}],\label{eq:bb}\\
&F^Q_{box}(x,y,z)
=
\frac{5 m_Q f_Q^2}{24}\left(\frac{m_Q}{m_{q_1}+m_{q_2}}\right)^2
\int \frac{\delta(1-a-b-c-d)dadbdcdd}{[a m_x^2+b m_y^2+(c+d) m_z^2]^2},
\end{align}
where $(q_1,q_2)$ are respectively $(d,s)$ for $K$,  $(b,d)$ for $B$, and  $(u,c)$ for $D$. Each of the last inequalities in Eqs.(\ref{eq:kk} -- \ref{eq:bb})
represents the upper bound on the corresponding experimental 
value \cite{pdg}.  Here we used
$f_K\approx0.156$ GeV, $f_B\approx0.191$ GeV, $m_K\approx0.498$ GeV,
and $m_B\approx5.280$ GeV.
\footnote{Since we assume that one of the neutrino masses to be zero with
normal ordering that leads to the 
{first column in $g$ to be almost zero, i.e.,
$(g)_{11,12,13}\approx 0$,} and so these constraints can easily be evaded. }
}

{\it Dark Matter}:
Here we identify $\eta_R$ as the DM candidate, and denote its mass by $m_{R}\equiv M_X$. 
{
{\it Direct detection}: We have a Higgs portal contribution to the DM-nucleon scattering process, which is constrained by direct detection search such as the LUX experiment~\cite{Akerib:2016vxi}.
Its spin independent cross section is simply given by~\cite{Baek:2016kud}
\begin{align}
\sigma_N\approx 2.12\times 10^{-42}\times\left(\frac{(\lambda_3+\lambda_4+2\lambda_5)v}{M_X}\right)^2 [{\rm cm}]^2,
\end{align}
where $\lambda_3$ and $\lambda_4$ are quartic couplings proportional to $(\eta^\dag\eta)(\Phi^\dag\Phi)$ and $(\eta^\dag\Phi)(\Phi^\dag\eta)$, respectively.
The current experimental minimal bound is $\sigma_N\lesssim 2\times 10^{-46}$ cm$^2$ at $M_X\approx 50$ GeV. Once we apply this bound on our model, we obtain $\lambda_3+\lambda_4+2\lambda_5\lesssim 2\times10^{-3}$. Hence we assume that all the Higgs couplings are small enough to satisfy the constraint, and we neglect DM annihilation modes via Higgs portal in estimating the relic density below. 
Notice here that photon and Z boson {fields transform as $V_\mu \to - V^\mu$ under 
charge-parity ($CP$) conjugation, while $X$ is $CP$-even.
Thus $X-X-\gamma(Z)$ couplings are not allowed because they violates the 
$CP$ invariance.}
\\
{\it Relic density}:}
We consider parameter region in which the DM annihilation cross section is $d$-wave dominant and the dark matter
particles annihilate into a pair of charged-leptons, via the process
$\eta_R \eta_R \to \ell_i \bar \ell_j$ with an $E_a$ exchange. 
{
Notice that there exist annihilation modes such as $\eta_R \eta_R \to W^+ W^-/ 2Z$ {arising} from the kinetic term.  
These modes require {a DM mass heavier} than at least 
$500$ GeV~\cite{Hambye:2009pw} in order to obtain the correct relic density where coannihilation processes should be included.  
This case is, however, not in favor of explaining the muon $g-2$ anomaly. {\it Thus we assume {that} $M_X\lesssim 80$ GeV} and $\eta_R \eta_R \to W^+ W^-/ 2Z$ processes are not kinematically allowed.
\footnote{
Here we impose the condition 
$m_Z/2(\approx 41$GeV$)\lesssim m_R+m_I$ to forbid the invisible decay of 
$Z$ boson in our numerical analysis, although the invisible decay of 
SM Higgs is automatically suppressed in the limit of zero couplings in the
Higgs potential. }}
Then the relic density is simply given by
\begin{align}
\Omega h^2\approx \frac{1.70\times10^{7} x_f^3}
{\sqrt{g^*} M_P d_{\rm eff} [{\rm GeV}]},
\quad
{ d_{\rm eff}\approx \sum_{(i,j,a)=1}^3\frac{|r_{ia}r^\dag_{a,j}|^2 M_X^6}{120\pi(M_{E_a}^2+M_X^2)^4},}
\end{align}
where $g^*\approx100$, $M_P\approx 1.22\times 10^{19}$, $x_f\approx25$.
In our numerical analysis below, we use the current experimental 
range for the relic density: $0.11\le \Omega h^2\le 0.13$~\cite{Ade:2013zuv}.\\

\section{Numerical analysis \label{sec:numerical}}
\begin{figure}[tb]
\begin{center}
\includegraphics[width=65mm]{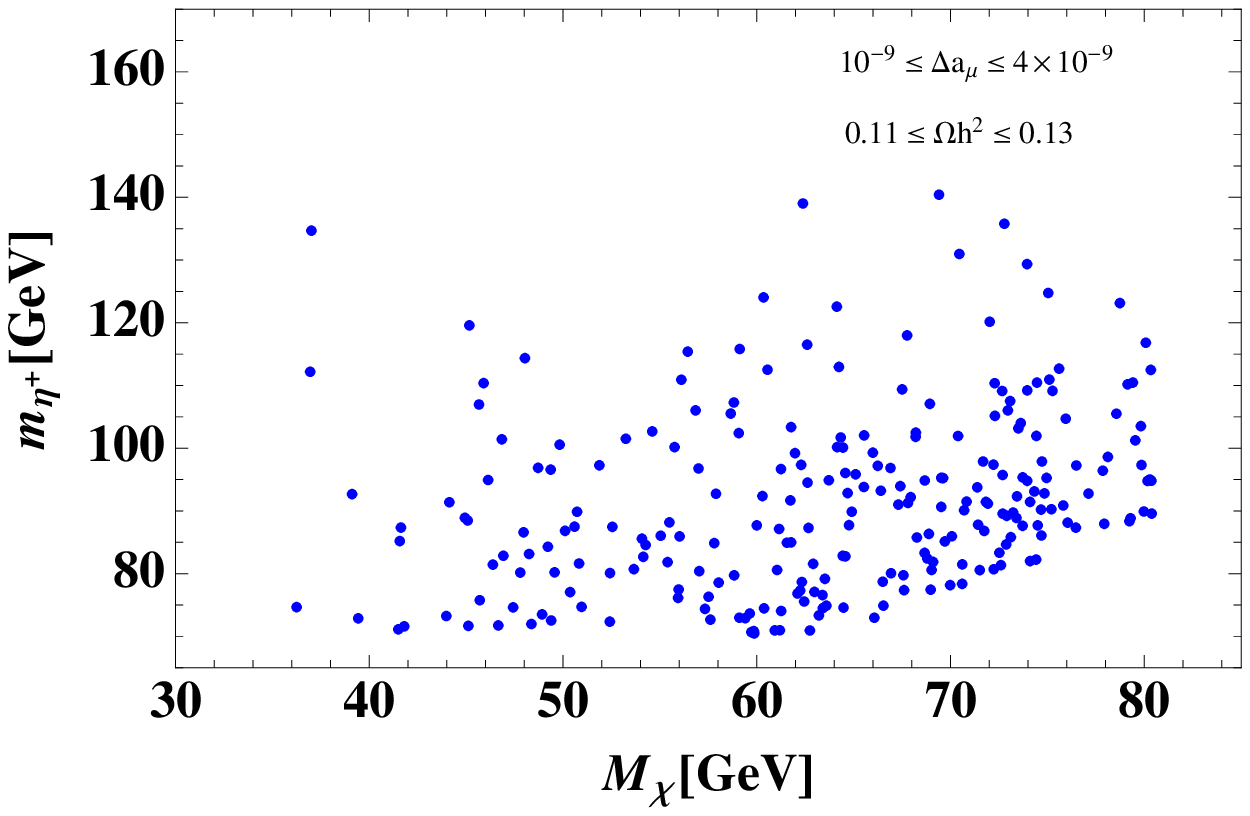}
\includegraphics[width=65mm]{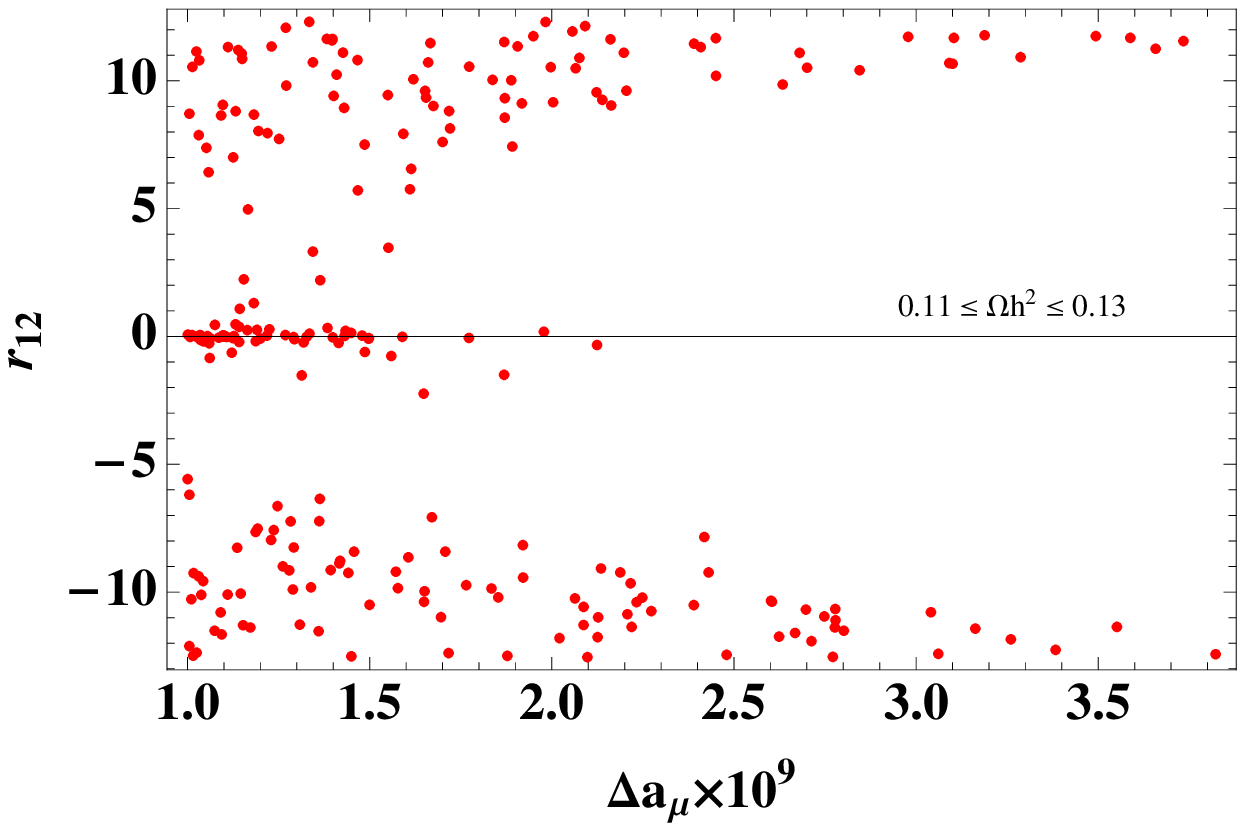}
\caption{
Scattering {\it} plots of the allowed parameter space sets to satisfy LFVs, oblique parameters, and invisible decay of $Z$ boson,
in the plane of $M_X$-$m_{\eta^\pm}$ in the left
panel; and in the plane of $\Delta a_\mu$-$r_{12}$ in the right panel,
where $r_{12}$ is the most relevant parameter to obtain the sizable muon $g-2$ and relic density of DM.
}
\label{fig:nums-1}
\end{center}
\end{figure}

As a first step, we perform the numerical analysis on the $r$ term
since this term is independent of the other parameters.
We prepare {\it 25 million} random sampling points
for the relevant input parameters as follows: 
{\begin{align}
& M_X \in [1, {80}]\text{GeV},\
m_{I}\approx m_{\eta^\pm} \in [1.1 \times M_X\,, {{200}}\,]\text{GeV},
\nn\\&
M_{1} \in [1.1\times M_X\,, {330}\,]\text{GeV},\
M_{2}  \in [M_1\,, 600\,]\text{GeV},
\
M_{3}  \in [M_{2}\,, 700\,]\text{GeV},\nn\\ &
r'\in[-9,\ln(4\pi)],
\label{range_scanning}
\end{align}
}
where we define {$M_i\equiv M_{E_i}=M_{N_i}$($i=1-3$)}, $r_{ij}\equiv (\pm1)\times10^{r'_{ij}}$ and the lower mass 
bound $1.1\times M_X$ is expected to forbid the coannihilation processes. 
Under such parameter ranges, we have found 246 allowed points, which are
shown in Fig.~\ref{fig:nums-1} satisfying all the constraints including 
LFVs, oblique parameters, and invisible decay of $Z$ boson, as discussed before.
The left panel shows the allowed region to be
\begin{align}
25  {\rm [GeV]} \lesssim M_X \lesssim 80 {\rm [GeV]},\quad  70  {\rm [GeV]}\lesssim m_{\eta^\pm} \lesssim 140 {\rm [GeV]},
\end{align}
where the lower bound of $\eta^\pm$ comes from the 
LEP experiment~\cite{Osland:2013sla, Akeroyd:2016ymd}.
The right panel shows the correlation of $r_{12}$ versus $\Delta a_\mu$,
where $r_{12}$ is the most relevant parameter to obtain the sizable 
muon $g-2$ and correct relic density of the DM.
It suggests that a rather small $r_{12}$ is possible to achieve the range 
$10^{-9}\le \Delta a_\mu \le 2\times 10^{-9}$, but we need a large $r_{12}$ 
for the range $2\times10^{-9}< \Delta a_\mu \le 2 - 4\times 10^{-9}$.
~\footnote{In addition to $r_{12}$, a little bit larger $r_{32}$ is also needed.}

\begin{figure}[tb]
\begin{center}
\includegraphics[width=65mm]{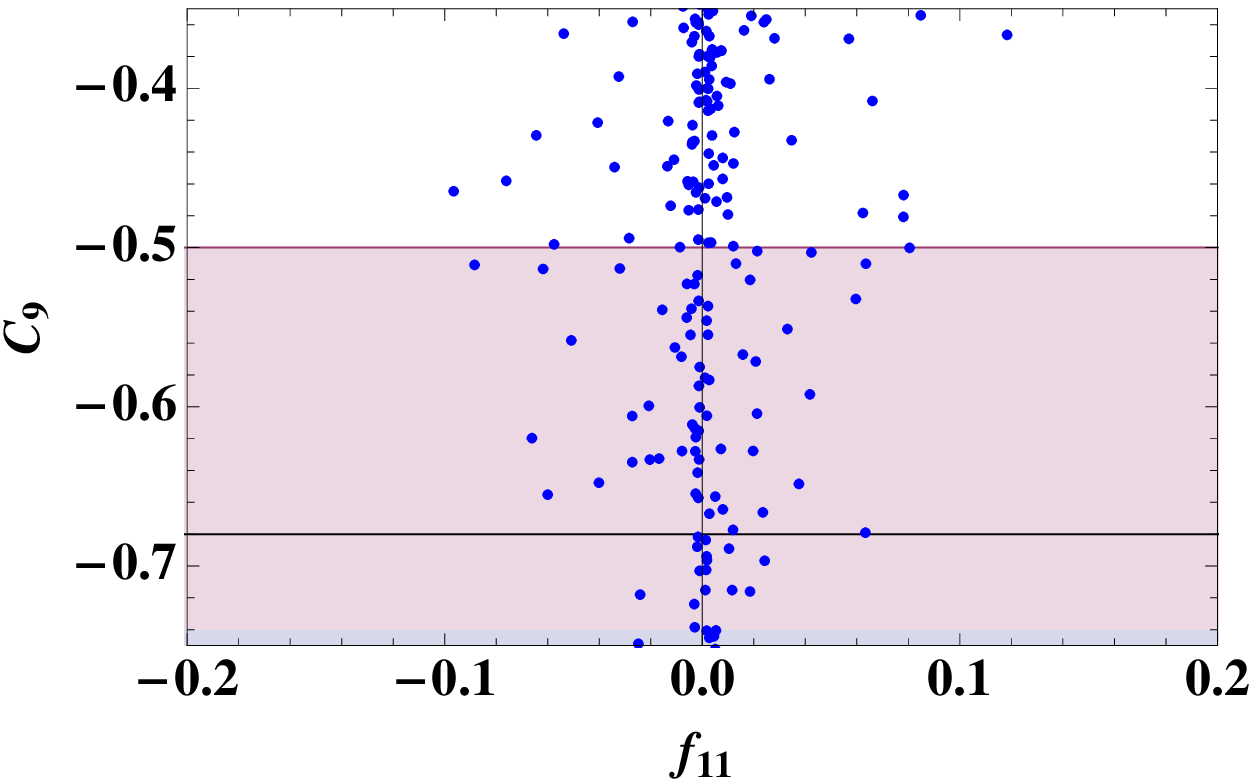}
\includegraphics[width=65mm]{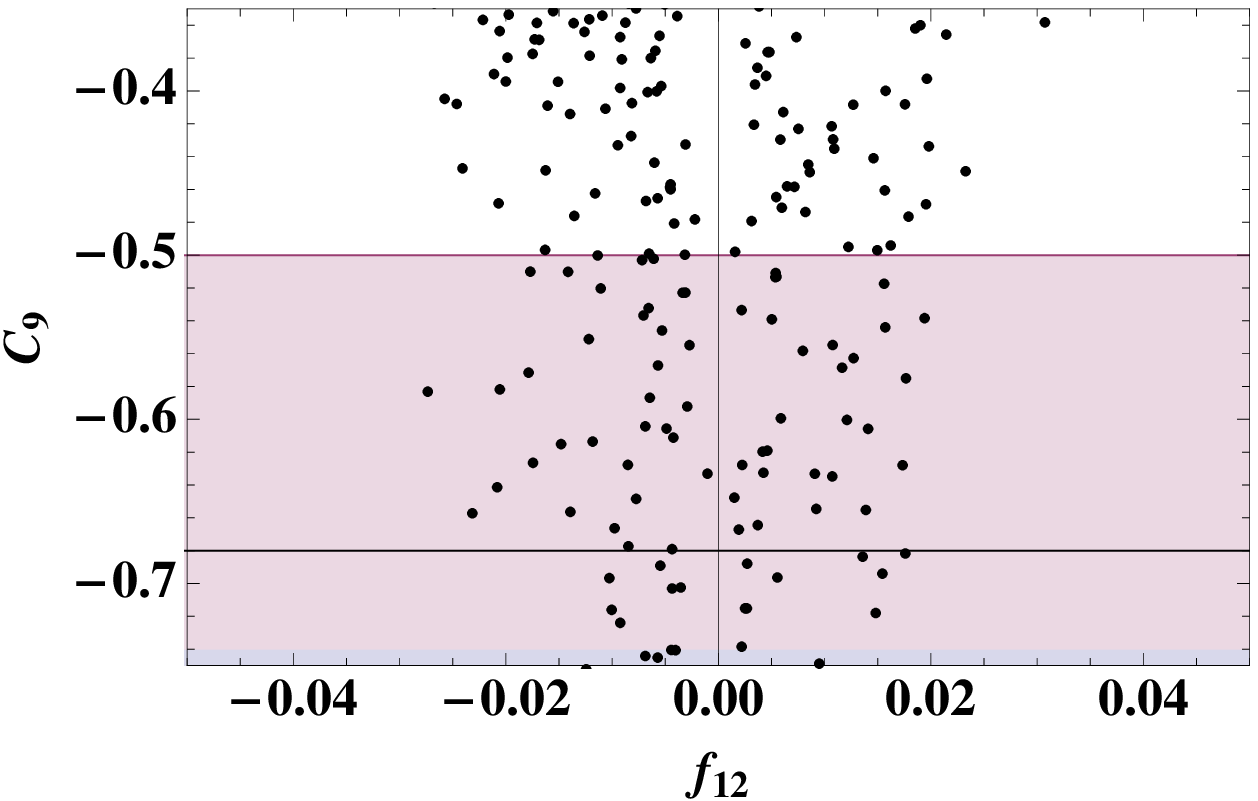}
\includegraphics[width=65mm]{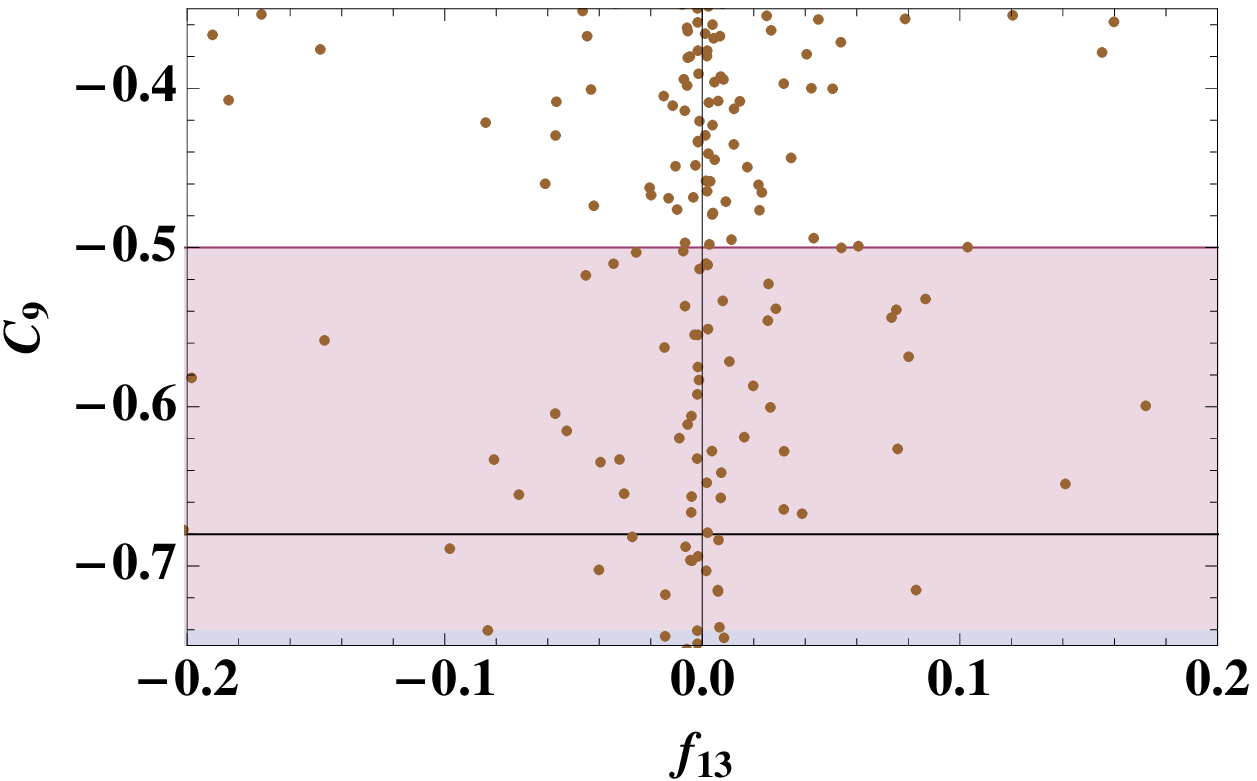}
\caption{
Scattering {\it} plots of the allowed parameter space sets
in the plane of $(f_{1j})(j=1-3)$-$C_{9}$, where the horizontal line is the best fit value of $C_9=-0.68$.
While  the thin red region $ [-0.85,-0.50]$ is the one at $1\sigma$ range.
Notice here that  we have taken the range $C_9\in [-0.75,-0.50]$ to satisfy $R_K$ at 1$\sigma$ confidential level~\cite{Descotes-Genon:2015uva}.
}
\label{fig:nums-2}
\end{center}
\end{figure}

In the next step, we attempt to find the parameter space points that can
also solve the $B\to K^{(*)}\mu\mu$ anomaly by contributing to the $C_{9(10)}$,
and at the same time satisfy all the constraints of the LFVs and FCNCs.
Since the number of parameters is getting more and more, we are content
with a benchmark point that we obtained in the first step.
We prepare the benchmark point for the masses to fix the 
three-loop neutrino function $F_{III}$
\footnote{This is technically difficult to obtain the whole numerical values,
 due to its complicated structure.} as follows:
\begin{align}
&M_X\approx 54.26\ [{\rm GeV}],\quad m_{\eta_I}\approx m_{\eta^\pm}\approx 84.57\ [{\rm GeV}],\\ 
& M_1\approx 277\ [{\rm GeV}],\quad M_2\approx 296\ [{\rm GeV}],\quad M_3\approx 401\ [{\rm GeV}],\\
&m_{\Delta_1}\approx1\ [{\rm TeV}],\quad m_{\Delta_2}\approx1.1\ [{\rm TeV}] ,\quad  m_{\Delta_3}\approx1.2\ [{\rm TeV}], 
\end{align}
where the above first two lines are provided by the first step 
so that $\Omega h^2\approx 0.120$ and $\Delta a_\mu\approx 2.67\times 10^{-9}$
are obtained with $r_{12}\approx -11.6$. The values in the last line 
are simply taken to evade the collider bound.
{
To satisfy bound on the direct detection search, $\lambda_3+\lambda_4+2\lambda_5\lesssim0.002$ is needed, where experimental upper bound is $\sigma_I\lesssim 2.2\times10^{-46}$ cm$^2$, while $\lambda_5\approx 0.035$ that comes from $m_{\eta_I}^2-M_X^2=2\lambda_5 v$. Therefore a little fine tuning is needed among $\lambda_3,\lambda_4,\lambda_5$.
}
With this benchmark point, we have the following values:
\begin{align}
& F_{III}[x_{1}]\approx-108.68,\quad
F_{III}[ x_{2}]\approx-287.73,\quad
 F_{III}[ x_{3}]\approx-317.11.
\end{align}
Also, we fix $-\lambda_0\lambda'_0 v^2/2\approx 1[{\rm GeV}^2]$ for simplicity. 
Then, we prepare {{\it 0.1 billion}} random sampling points
for the following relevant input parameters:
{\begin{align}
& [a_{12},a_{13},a_{23}] \in [-1-i, 1+i]\times [10^{-3},1],\quad
f'\in[-3,\ln(4\pi)],\quad h'\in[-3,\ln(4\pi)],
\end{align}
}
where we define $f(h)_{ij}\equiv (\pm1)\times10^{f'(h')_{ij}}$. 
In these parameter ranges, we have found {{\it 870}} allowed 
points shown in Fig.~\ref{fig:nums-2}, which satisfy all the constraints 
as discussed before. If we focus on the best-fit value of $C_9$, 
these figures show that the Yukawa couplings $f_{1j}(j=1-3)$ are 
very restricted as 
\begin{align}
|f_{11}| \lesssim {0.1},\quad  |f_{12}| \lesssim {0.02},\quad  |f_{13}| \lesssim {0.2},
\end{align}
where these coupling may affect the collider physics.

{\it Collider Phenomenology}: There are two types of new particles in this 
model other than the SM particles: leptoquarks $\Delta_{1,2,3}^a$ and
a set of scalar bosons  $\eta^{\pm,0}$ resulting from an isodoublet scalar field.

Note that $\Delta_{1,2}^a$ are assigned with $Z_2=-1$ while
$\Delta_{3}^a$ are assigned with $Z_2=+1$. All $\Delta_{1,2,3}$ can be
pair produced at hadronic colliders and being searched at the LHC. 
The direct search bound is roughly 1 TeV \cite{lhc-lq}. 
On the other hand, only $\Delta_3^a$ can participate in the
 4-fermion contact interaction, because of the $Z_2$ parity. 
The bound from the 4-fermion contact interaction was worked out in 
Ref.~\cite{Cheung:2016frv,Cheung:2016fjo} that the bound is currently
inferior to the direct search bound of about 1 TeV.  Therefore, we shall
use 1 TeV as the current bound on the $\Delta_{1,2,3}$ bosons.

The isodoublet field $\eta$ gives rise to a pair of  charged bosons
$\eta^\pm$, a scalar $\eta_R$, and a pseudoscalar $\eta_I$.  The charged 
bosons $\eta^\pm$ can run in the triangular loop vertex of $H\gamma\gamma$.
Nevertheless, we can suppress such effects by choosing the $\lambda_5$ term
in Eq.~(\ref{eq:Yukawa}) very small. As we have explained above such a term
is small to avoid the conflict of the direct detection bound.
Although the interaction between the $\eta$ field and the SM Higgs field is 
suppressed for the above reasons, the $\eta$ can still interact through
the kinetic term as it has $SU(2)_L$ and $U(1)_Y$ interactions. 
We expect some typical interactions with the gauge bosons:
\[
  Z \eta^+ \eta^-, \;\; Z \eta_R \eta_I, \;\; W^+ \eta^- \eta_R, \;\;
{\rm etc}\;.
\]
An interesting signature would be Drell-Yan type production
of $\eta^\pm \eta_R$ via a virtual $W$. The $\eta^\pm$ decays into 
multi-leptons and $\eta_R$ via the virtual $L'$ fields. Therefore, 
the final state consists of multi-charged-leptons and missing energies.
Similarly, in the process $pp \to Z^* \to \eta_R \eta_I$, the $\eta_I$
would decay into $\eta_R$ eventually with a number of very soft leptons,
which may not be detectable. Therefore, the best would be the one 
produced via virtual $W$.

\section{Conclusions}
We have investigated a three-loop neutrino mass model 
with some leptoquark scalars with $SU(2)_L$-triplet, in which we have explained the anomaly in $B\to K^{(*)} \mu^+\mu^-$, sizable muon $g-2$, bosonic dark matter, satisfying all the constraints such as LFVs, FCNCs, invisible decay, and so on.
Then we have performed the global numerical analysis and shown the allowed region, in which we have found restricted parameter space, {\it e.g.,}
\begin{align*}
& 25  {\rm [GeV]} \lesssim M_X \lesssim 80 {\rm [GeV]},\quad  70  {\rm [GeV]}\lesssim m_{\eta^\pm} \lesssim 140 {\rm [GeV]},
\\
& |f_{11}| \lesssim {0.1},\quad  |f_{12}| \lesssim {0.02},\quad  |f_{13}| \lesssim {0.2}.
\end{align*}  
We find that $\sim O(100)$ GeV inert doublet scalar is preferred to obtain sizable muon $g-2$.
Thus these light inert scalars could be tested by collider experiments such as LHC in which these scalars are produced via electroweak processes.
The promising signature of our model comes from the process $pp \to W^\pm \to \eta^\pm \eta_R$ which provides signals of multi-leptons plus missing transverse momentum.

\section*{Acknowledgments}
This work was supported by the Ministry of Science and Technology
of Taiwan under Grants No. MOST-105-2112-M-007-028-MY3.

\begin{appendix}
\section{Loop function}
Here we show the explicit form of three-loop function $F_{III}$, which is given by
\begin{align*}
&F_{III}=\int [dx_i]\frac{\delta(\sum_{i=1}^5 x_i-1)(x_2-1)^2}{x_1^2}
\int \frac{[dy_i] \delta(\sum_{i=1}^3 y_i-1)}{[(1+y_1)x_1x_2-x_1-x_2+1]^3}
\int \frac{[dz_i] \delta(\sum_{i=1}^3 z_i-1)z_1^2}{[y_1\Delta_3+z_2 r_{\delta_{13}^{(2)}} +z_3 r_{d_a}]^3},\\
&
\Delta_3=\frac{x_2-1}{x_1[(1+y_1)x_1x_2-x_1-x_2+1] }
\left( x_1 r_{\delta_{13}^{(1)}} +x_2r_{\delta_{13}^{(2)}}+x_3r_{b}+x_4 r_{\eta_R}+x_5 r_{\eta_I}\right)\nn\\
&-\frac{x_1x_2}{A^2 y_1[(1+y_1)x_1x_2-x_1-x_2+1]}
\left( y_2 r_{\delta_{13}^{(3)}}+y_3r_{d_c}\right),
\end{align*}
where $A\equiv x_1/(x_2-1)$, $[dx_i]\equiv \int_0^1dx_1\int_0^{1-x_1}dx_2\int_0^{1-x_1-x_2}dx_3\int_0^{1-x_1-x_2-x_3}dx_4$, $[dy_i]\equiv \int_0^1dy_1\int_0^{1-y_1}dy_2$, and $[dz_i]\equiv \int_0^1dz_1\int_0^{1-z_1}dz_2$.

\end{appendix}

\end{document}